\documentclass[12pt, twocolumn]{article}
\usepackage{amsmath,latexsym}
\usepackage{graphicx}
\begin{document}

 \title{\Huge Can electro-magnetic field, anisotropic  source and varying $\Lambda$ be
 sufficient  to produce wormhole spacetime ?
  }
 \author{F.Rahaman$^*$, M.Kalam$^{\dag}$
 and K A Rahman$^*$}

\date{}
 \maketitle
 \begin{abstract}
It is well known that solutions of general relativity which allow
for traversable wormholes require the existence of exotic matter (
matter that violates  weak or null energy conditions [WEC or NEC]
).  In this article,  we provide a class of exact solution for
Einstein-Maxwell field equations describing  wormholes assuming
the erstwhile cosmological term $\Lambda$ to  be space variable ,
viz., $\Lambda = \Lambda  (r)$.
 The source considered here not only a matter entirely but a sum of matters i.e.
anisotropic matter distribution, electromagnetic field and
cosmological constant whose effective parts obey all energy
conditions out side the wormhole throat. Here violation of energy
conditions can be compensated by varying cosmological constant.
The important feature of this article is that one can get wormhole
structure, at least theoretically, comprising with physically
acceptable matters.

\end{abstract}

  \footnotetext{
  Pacs Nos :  04.20 Gz,04.50 + h, 04.20 Jb


 $*$Dept.of Mathematics, Jadavpur University, Kolkata-700 032, India

                                  E-Mail:farook\_rahaman@yahoo.com\\
$\dag$Dept. of Phys. , Netaji Nagar College for Women, Regent Estate, Kolkata-700092, India.\\
E-Mail:mehedikalam@yahoo.co.in\\
}
    \mbox{} \hspace{.2in}

\title{ \underline{\textbf{Introduction}}: }

We know a wormhole is a hypothetical  topological feature of
spacetime that  connects two distinct spacetimes. The wormhole
idea comes from  Einstein's theory of general  relativity [1].  It
is the solution  of Einstein equation shared by the violation of
null energy condition. The matter  that characterized above stress
energy tensor is known as exotic matter. Needless to say, the
notion of this exotic matter is bizarre. In  spite of, several
physicists have constructed wormholes by assuming different forms
of exotic matter. Sushkov[2], Lobo[3], Kuhfittig[4],
Zaslovskii[5], Rahaman et al[6] have presented wormhole solutions
comprising of phantom energy. Lobo [7] , Rahaman et al [7] and
Rashid et al [7] have shown that wormholes may be supported by the
Chaplygin gas. Das et al[8] have studied wormhole with  Tachyonic
field. Mansouryar[9] and Khabibullin A et al [10] have  assumed
Casimir field for exotic matter source. Rahaman et al [11] have
studied wormhole  in presence of C-field. Also Rahaman et al [12]
have shown that wormholes may exist in  Kalb-Ramond spacetime.

To avoid this bizarre form of matter distribution, several authors
used scalar tensor theory of gravity to construct wormholes[13].
Though Visser et al[14] showed and latter supported by
Kuhffitig[15], Nandi et al[16] and Fewster et al[17] that the
amount of exotic matter needed can be made arbitrarily small for
constructing wormholes but  no matter how difficult to deal with
exotic matter. So  we are trying to provide a  prescription how to
get a wormhole comprising with physically acceptable matters. We
give a class of solution of Einstein-Maxwell field equations
describing  wormholes assuming cosmological term $\Lambda$ to  be
space variable. The source considered here not only a matter
entirely but a sum of matters i.e. anisotropic matter
distribution, electromagnetic field and cosmological constant
whose effective parts obey all energy conditions out side the
wormhole throat. Here violation of energy conditions can be
compensated by varying cosmological constant. The assumption of
variable $\Lambda$ is not new [  see ref.[18], for review  ].
Several authors have discussed the contribution of space
dependence $\Lambda$ to the effective gravitational mass of the
astrophysical systems[19]. The solutions of Einstein field
equations with variable $\Lambda$ have a wider range of
application to discuss more accurately the local massive objects
like galaxies[20] and energy density of classical electron[21].
So, it is not unnatural to  inclusion of $\Lambda$ on an
anisotropic static spherically symmetric  source to construct
wormholes. Recently Lemos et al [22] have studied extensively
wormhole geometry in presence of $\Lambda$ where $\Lambda$ is a
constant. The aim of the present investigation is to construct
stable traversable wormhole with realistic matter sources.
\\
\\

\title{ \underline{\textbf{Basic equations for constructing wormholes}}: }

Let  us consider  a static, spherically symmetric matter
distribution corresponding to the line element
\begin{equation}
               ds^2=  - e^{\nu(r)} dt^2+ e^{\mu(r)} dr^2+r^2( d\theta^2+sin^2\theta
               d\phi^2)
         \label{Eq3}
          \end{equation}
          The Einstein-Maxwell field equations for the above spherically
          symmetric metric corresponding to the charged
          anisotropic matter distribution in presence of varying
          $\Lambda$ are given by
\begin{equation}e^{-\mu}
[\frac{\mu^\prime}{r} - \frac{1}{r^2} ]+\frac{1}{r^2}= 8\pi \rho +
E^2 + \Lambda
\end{equation}
\begin{equation}e^{-\mu}
[\frac{1}{r^2}+\frac{\nu^\prime}{r}]-\frac{1}{r^2}= 8\pi p_r - E^2
- \Lambda \end{equation}
\begin{equation}\frac{1}{2}e^{-\mu}
[\frac{1}{2}(\nu^\prime)^2+ \nu^{\prime\prime}
-\frac{1}{2}\mu^\prime\nu^\prime + \frac{1}{r}({\nu^\prime-
\mu^\prime})] =8\pi p_t + E^2 - \Lambda\end{equation} and
\begin{equation}(r^2E)^\prime = 4\pi r^2 \sigma e^{\frac{\mu}{2}}
\end{equation}
Equation (5) can equivalently be expressed in the form
\begin{equation} E(r) = \frac{1}{r^2}\int_0^r 4\pi r^2 \sigma
e^{\frac{\mu}{2}}dr = \frac{q(r)}{r^2}
\end{equation}
where $q(r)$ is the total charge of the sphere under
consideration. Also, the conservation equation is given by
\begin{equation} \frac{d}{dr} ( p_r - \frac {\Lambda}{ 8 \pi })
+ ( \rho + p_r ) \frac{ \nu^\prime }{2} = \frac{1}{ 8 \pi
r^4}\frac{dq^2}{dr} + \frac{2(p_t - p_r)}{r}
\end{equation}
Here, $ \rho, p_r, p_t, E, \sigma$   and  $ q $ are respectively
the matter energy density, radial and  tangential pressures,
electric field strength, electric  charged  density and electric
charge. The prime denotes derivative with respect to 'r'.
\\
\\
\\
\\
\title{ \underline{\textbf{Solutions}}: }

Now to get exact solutions, we assume the following assumptions:

(a) \begin{equation} \nu(r) = 0 \end{equation} \textbf{Argument:}
One of the traversability properties is the tidal gravitational
forces experienced by a traveller must be reasonably small. So, we
assume a zero tidal force as seen by the stationary observer. Thus
one of the traversability conditions is automatically satisfied.

(b) \begin{equation}  p_t = n p_r \end{equation}

\textbf{Argument:} Pressures are anisotropic with $ p_t < p_r $.

(c)  \begin{equation}  p_r = m \rho \end{equation}
\textbf{Argument:} The above equation indicates the equation of
state with $ 0 < m < 1 $.

 (d)   $\frac {\Lambda}{ 8 \pi } \propto  p_r      $

i.e.
 \begin{equation}     \frac {\Lambda}{ 8 \pi }  = a p_r \end{equation}

 ( $a$ is proportional constant  )

  \textbf{Argument:} The vacuum energy ( which is equivalent to $ \Lambda $
 ) can be  thought as a contribution of the  energy  stress
 components.

(e) \begin{equation}  \sigma e^{\frac{\mu}{2}} = \sigma_0 r^s
\end{equation}

( $\sigma_0$  and  s  are  arbitrary constants  )

 \textbf{Argument:}
In usual sense, the term $  \sigma e^{\frac{\mu}{2}} $ occurring
inside the integral sign in the equation (6), is called the
volume charge density and hence the condition $  \sigma
e^{\frac{\mu}{2}} = \sigma_0 r^s $ , can equivalently be
interpreted as the  volume charge density being polynomial
function of 'r'. The constant $ \sigma_0  $ is the charge density
at $ r = 0 $, the center of the charged matter [19].

Taking into account of equations (8) - (12), one  gets the
following solutions of the field equations (2)  - (7) as

\begin{equation}  q^2(r)  =  \frac{16\pi^2\sigma_0^2}{(s+3)^2}
r^{2s+ 6}
\end{equation}

\begin{equation}  E^2(r)  =  \frac{16\pi^2\sigma_0^2}{(s+3)^2}
r^{2s+ 2}
\end{equation}

\begin{equation}  p_r  =  Dr^{\frac{-2(1-n)}{(1-a)}} + \frac{4\pi\sigma_0^2}{P}
r^{2s+ 2}
\end{equation}

where  $ P = (s+3)[2(1-n)+(2s+2)(1-a)]$ and  D is an integration
constant.

\begin{equation}  e^{-\mu}   =  1 - \frac{b(r)}{r}
\end{equation}

where,

\begin{equation}  b(r)  =  F r^{\frac{(2n +1 - 3a)}{(1-a)}} + X
r^{2s+ 5}
\end{equation}

where, $ F = \frac{8\pi D ( 1 - a )( a + \frac{1}{m})} {( 2n + 1
-3a )}$

 and

 $ X = \frac{16\pi^2\sigma_0^2}{(2s +  5 )}[
\frac{1}{(s+3)^2} + \frac{( a + \frac{1}{m})}{P}]$
 \pagebreak
 \\
\begin{figure}[htbp]
    \centering
        \includegraphics[scale=.6]{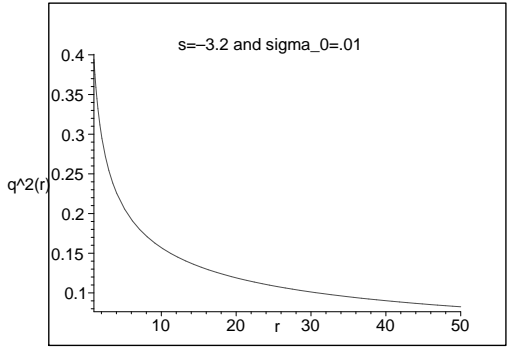}
        \caption{ Electric charge  with respect to radial coordinate 'r'. }
    \label{fig:1}
\end{figure}

\begin{figure}[htbp]
    \centering
        \includegraphics[scale=.6]{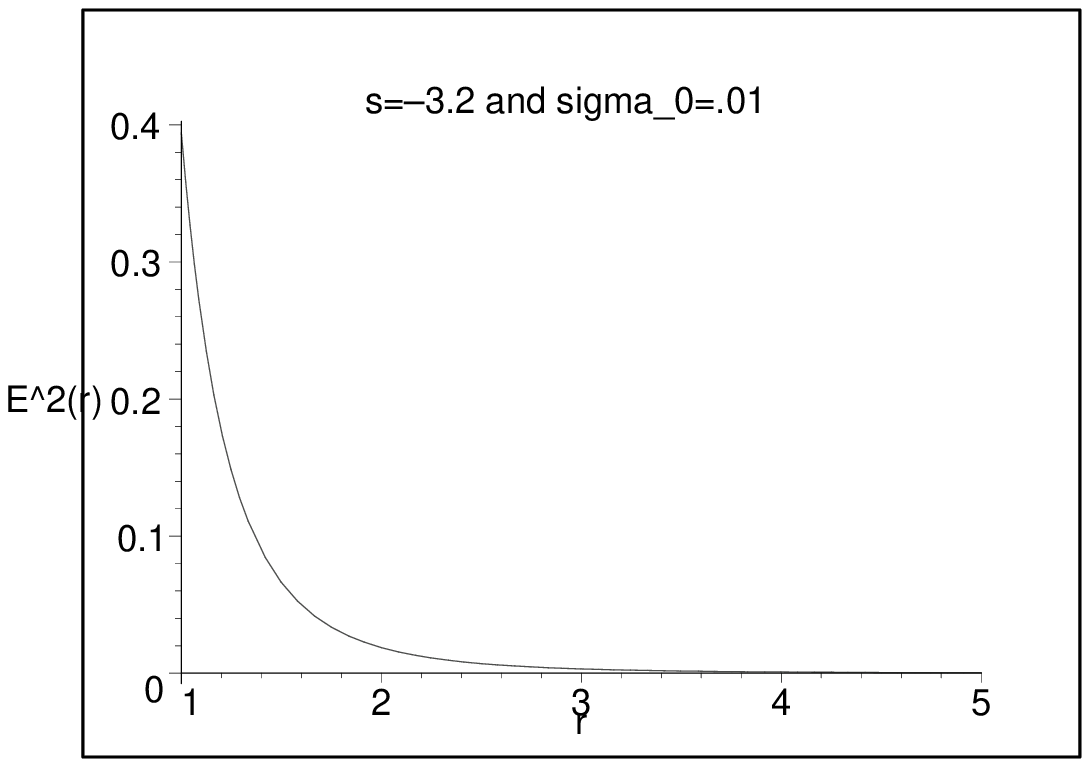}
        \caption{ Electric field strength  with respect to radial coordinate 'r'. }
    \label{fig:1}
\end{figure}

\begin{figure}[htbp]
    \centering
        \includegraphics[scale=.6]{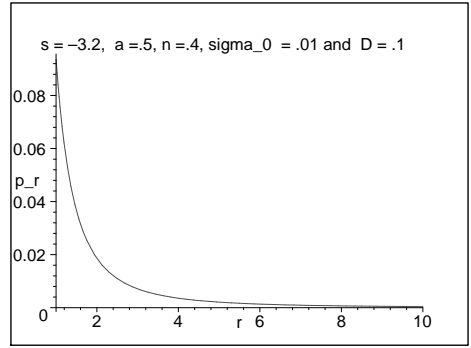}
        \caption{ Radial pressure  with respect to radial coordinate 'r'. }
    \label{fig:1}
\end{figure}

\begin{figure}[htbp]
    \centering
        \includegraphics[scale=.6]{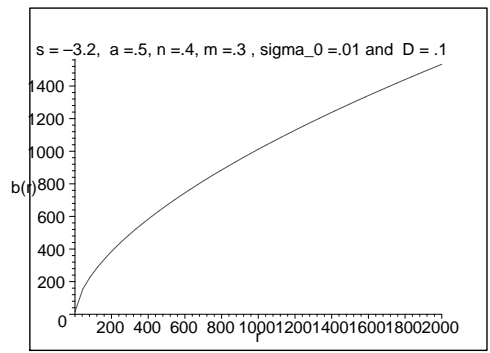}
        \caption{ Shape function  with respect to radial coordinate 'r'. }
    \label{fig:1}
\end{figure}
 \pagebreak
\title{ \underline{\textbf{Properties of the solutions}}: }
Since the space time is asymptotically flat i.e.
$\frac{b(r)}{r}\rightarrow 0 $ as $ r  \rightarrow \infty $, the
Eq.{(17)} is consistent only when $\frac{(2n+1-3a)}{(1-a)} - 1 < 0
$  and $ 2s  + 4 < 0  $.

These imply,
\begin{equation}   n < a
\end{equation}
and
\begin{equation}   s < -2
\end{equation}
Also, as $ \mid r \mid \rightarrow \infty $, $p_r$, $q^2(r)$ and
$E^2(r)$ $ \rightarrow 0 $, so one has to take the following
restriction on 's' as
\begin{equation}   s < -3
\end{equation}

Here the throat occurs at $ r= r_0 $ for which $ b(r_0) = r_0 $
i.e. $ 1 = F r_0^{\frac{(2(n - 1)}{(1-a)}} + X r_0^{2s+ 4}$. For
the suitable choices of the parameters, the graph of the function
$ G(r) = b (r) - r$ indicates the point $r_0$ , where G(r) cuts
the 'r' axis (see fig. 5 ).  From the graph, one can also note
that  when $r>r_0 $, $G(r)< 0$ i.e. $  b(r) -r < 0 $. This implies
$ \frac{b(r) }{r} < 1 $ when $r>r_0 $.

\begin{figure}[htbp]
    \centering
        \includegraphics[scale=.6]{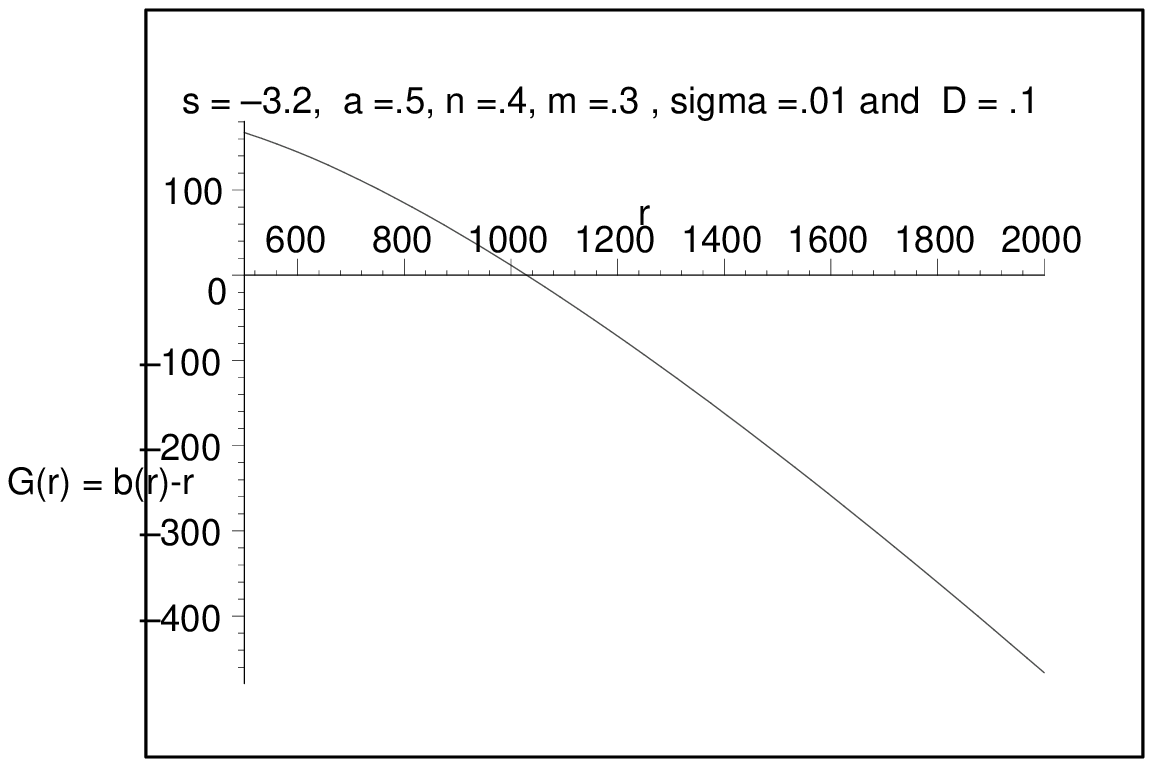}
        \caption{ Throat occurs where $G(r)$ cuts 'r' axis }
    \label{fig:1}
\end{figure}

Thus our solution describing a static spherically symmetric
wormhole supported by anisotropic matter distribution in presence
of electromagnetic field and varying $ \Lambda$.

\pagebreak

\title{ \textbf{\underline{Stability} }: }

To study the stability, we match our interior wormhole solution
to the exterior Reissner-Nordstr\"{o}m Black hole solution

$ds^2=  -(1 - \frac{2M}{r} + \frac{Q^2}{r^2}) dt^2 + \frac{dr^2}{
               (1 - \frac{2M}{r} + \frac{Q}{r^2})} + r^2
               d\Omega_2^2,$

at the junction interface S, situated outside the event horizon,
$ a > r_h = M \pm \sqrt{M^2 - Q^2 }$, one needs to use extrinsic
curvature or second fundamental forms associated with two sides
of the shell 'S' as  $ K_{ij}^\pm =  - n^\pm_{\mu;\nu} e^\mu
_{(i)}e^\nu _{(j)}$, where $n^\pm$ are the unit normals to S and
$e^\mu _{(i)}$ are the components of the holonomic basis vectors
tangent to S. Using the Darmois-Israel formalism, we write
Lanczos equations for the surface stress energy tensors $S_j^i$ at
the junction interface S as
\begin{equation}S_j^i  =  - \frac{1}{8\pi} ( [K_j^i]  - \delta_j^i K )\end{equation}
where
 $ S_j^i = diag ( - \sigma , p_{\theta}, p_{\phi}) $
 is the  surface  energy tensor with $\sigma$ , the surface density and $p_\theta$ and $p_\phi$,
  the
surface pressures and $[ K_{ij} ] = K_{ij}^+ - K_{ij}^-$ and $ K
= [K_i^i ] = trace [K_{ij}]$.

To analyze the dynamics of the wormhole, we permit the radius of
the throat to become a function of time, $ a \rightarrow a(\tau)$.
Now taking into account equation (21), one can find,
\begin{equation}
               \sigma =  - \frac{1}{4\pi a}[ \sqrt{1 - \frac{2M}{a} + \frac{Q^2}{a^2} +
               \dot{a}^2}- \sqrt{1 - Fa^u - Xa^w  +
               \dot{a}^2}]
               \end{equation}
$
              p_{\theta} = p_{\phi} =
  p =  \frac{1}{8\pi a}\frac{1 - \frac{M}{a} + a \ddot{a} +
               \dot{a}^2}{\sqrt{1 - \frac{2M}{a} + \frac{Q^2}{a^2} +
               \dot{a}^2}}\\ - \frac{(1 - Fa^u - Xa^w  +
               \dot{a}^2) + a \ddot{a} + \frac{\dot{a}^2(Fu a^u + Xw
               a^w)}{2(1 - Fa^u - Xa^w)}} {\sqrt{1 - Fa^u - Xa^w  +
               \dot{a}^2}}
              $
\begin{equation} \end{equation}
[ over dot  means  the derivatives with respect to $\tau$ and $ u
= \frac{2(n-a)}{(1-n)}$  ; $ w = 2s  + 4 $  ]

Using conservation identity \\ $ S^i_{j;i} = - [  \dot{\sigma} +
2\frac{\dot{a}}{a}( p + \sigma )] $ , one can get the following
expression as

\begin{equation} {\sigma^\prime} = -  \frac{2}{a}( p + \sigma )+ Y \end{equation}
 where,

$ Y =  - \frac{1}{4\pi a^2} [ \frac{(Fu a^u + Xw
               a^w)}{2(1 - Fa^u - Xa^w)}\times\\ \sqrt{1 - Fa^u - Xa^w  +
               \dot{a}^2}] $
\begin{equation} \end{equation}

 Rearranging equation (22), we obtain the thin shell's
equation of motion
            \begin{equation}  \dot{a}^2 + V(a)= 0  \end{equation}
                Here  the potential is defined  as
\begin{equation}
              V(a) =  \frac{1}{2} (f_1 + f_2)  - 4\pi^2a^2
            \sigma^2 - \frac{(f_1 - f_2)^2}{64\pi^2a^2
            \sigma^2}
                 \end{equation}
where,
\begin{equation}
            f_1 =   1 - \frac{2M}{a} + \frac{Q^2}{a^2}
           ;  f_2 = 1 - Fa^u - Xa^w
               \end{equation}
Linearizing around a static solution situated at $a_0$, one can
expand V(a) around $a_0$ to yield

$
              V =  V(a_0) + V^\prime(a_0) ( a - a_0) + \frac{1}{2} V^{\prime\prime}(a_0)
              ( a - a_0)^2 \\+ 0[( a - a_0)^3]$
      \begin{equation}           \end{equation}
where prime denotes derivative with respect to $a$.

Since we are linearizing around a static solution at $ a = a_0 $,
we have $ V(a_0) = 0 $ and $ V^\prime(a_0)= 0 $. The stable
equilibrium configurations correspond to the condition $
V^{\prime\prime}(a_0)> 0 $.\\ Now we define a parameter $\beta$,
which is interpreted as the speed of sound, by the relation
\begin{equation}
              \beta^2(\sigma) = \frac{ \partial p}{\partial
              \sigma}|_\sigma
                 \end{equation}
Using equation (24), we have
\begin{equation}
              \beta^2(\sigma) = -1 + \frac{ a}{2
              \sigma^\prime}[ \frac{2}{a^2} ( p + \sigma ) + Y^\prime
              -
              \sigma^{\prime\prime}]
                 \end{equation}
The wormhole solution is stable if $ V^{\prime\prime}(a_0)> 0 $
i.e.

 $
\frac{4\sigma^\prime}{a^3}[\frac{(f_1-f_2)^2}{32\pi^2\sigma^3a^2}-8\pi^2\sigma
a^2](1+ \beta_0^2) <
             \frac{1}{2}(f_1^{\prime\prime}+f_2^{\prime\prime})-
             8\pi^2(\sigma^2-4a\sigma\sigma^\prime-2a^2\sigma^{\prime
             2})
\linebreak
             +\frac{2}{a^2}[(p+\sigma)+Y^\prime][\frac{(f_1-f_2)^2}{32\pi^2\sigma^3a^2}-8\pi^2\sigma a^2]
              -\frac{(f_1^\prime-f_2^\prime)^2}{32\pi^2\sigma^2 a^2}
              -\frac{(f_1-f_2)(f_1^{\prime\prime}-f_2^{\prime\prime})}{32\pi^2\sigma^2 a^2}
\linebreak
              +\frac{(f_1-f_2)(f_1^\prime-f_2^\prime)}{16\pi^2\sigma^2 a^2}(\frac{\sigma^\prime}{\sigma}+\frac{\sigma^\prime}{a}+\frac{2}{a})
              -\frac{(f_1-f_2)^2}{16\pi^2\sigma^2
              a^2}(\frac{2\sigma^\prime}{a\sigma}+\frac{3}{2}(\frac{\sigma^\prime}{\sigma})^2+\frac{3}{2a^2})$

or,
\begin{equation}
              \beta_0^2 < \frac{ A - B + C - S - T  + G - H}{N - L} - 1
                 \end{equation}
                 where A, B, C, S, T, G, H, N, L are given
                 in the appendix at $ a = a_0$.

Thus if one treats $a_0$, M and Q and other parameters   are
specified quantities, then the stability of the configuration
requires the above restriction on the parameter $\beta_0$. This
means there exists some part of the parameter space where the
throat location is stable.  $ $   [ To get geometrical
information, one can show the stability region graphically by
plotting $ \beta_{\mid (a=a_0)}$ vs. $x= \frac{M}{a_0}$ and taking
all other parameters as  known quantities. The stability region is
given below the curve.  ]

\pagebreak

\title{ \underline{\textbf{Traversability conditions}}: }

Now we will focus on the usability of our wormhole structure i.e.
to check whether it is useful  for the travellers of modern
civilizations. To travel through a wormhole, the tidal
gravitational forces experienced by a traveller must be reasonably
small. According to Morris and Thorne [1], the acceleration felt
by the traveller should not exceed Earth's gravity. Thus the tidal
accelerations between two parts  of the traveller's body,
separated by say, 2 meters, must less than the gravitational
acceleration at Earth's surface $g_{earth}$ ( $ g_{earth}\approx
10 m / sec^2$  ). Due to Morris and Thorne [1], the testing
tangential tidal constraint is given by ( assuming $ \nu ^ \prime
= 0 $ )

$ |R_{t \theta t \theta}| = R_{t \phi t \phi}| =
|\frac{\beta^2}{2r^2} (\frac{v}{c})^2 ( b^\prime - \frac{b}{r} ) |
\leq \frac{g_{earth}}{2c^2m } \approx \frac{1}{10^{10}  m^2} $

with $ \beta = \frac{1}{\sqrt{1-(\frac{v}{c})^2}}$ and c is the
velocity of light.

[ The above inequality indicates a restriction on traveller's
velocity $v(r)$ with which the traveller crosses the wormhole ]

For $ v << c $, we have $\beta  \approx  1 $ and substituting the
expression of $ b(r) $, we get

$ \frac{v}{c} < \frac{1}{10^8} \sqrt{\frac{1}{\frac{Fu}{2} r^{u-2}
+ \frac{Xw}{2} r^{w-2}}}$

The above inequality represents the tangential tidal force and
restrict the speed v of the while crossing the wormhole. Here
radial acceleration is zero since $ R_{rtrt} = 0 $, for our
wormhole spacetime. Acceleration felt by a traveller should less
than the gravitational acceleration at earth surface, $g_{earth}$.
The condition imposed by Morris and Thorne [1] as

$ |\textbf{f}| = |\sqrt{[ 1 - \frac{b(r)}{r}}] \beta ^ \prime c^2|
\leq  g_{earth}
 $ [ for $
\nu^\prime = 0 $]

 For the traveller's velocity $ v = constant$, one finds that $
 |\textbf{f}|= 0
 $.
 In our model the the above condition  is automatically satisfied, the
 traveller feels a zero gravitational acceleration.

\title{ \underline{\textbf{Final Remarks}}: }

Our  aim in this article is to search reasonable matters that
produce wormhole like spacetime. We have been able to show that if
we are supplied anisotropic matter source and electromagnetic
field along with varying $\Lambda$, then one could construct, at
least theoretically, a stable traversable wormhole. One can note
that $ \rho_{effective} > 0, $ $ \rho_{effective}$ + $p_{r}$
$_{effective}> 0 $, $ \rho_{effective}$ + $p_{t}$ $_{effective}> 0
$ for all  $ r > r_0 $ i.e. all energy conditions are satisfied
out side the throat. But at the throat i.e. at $r = r_0$, NEC is
violated. Nevertheless this wormhole has been constructed nearly
accessible matter sources.

The collections of anisotropic matter and electromagnetic field
are not difficult. The only difficult task is to collect the
source '$\Lambda$'. According to Zeldovich[23], $\Lambda$ is
nothing but the vacuum energy density due to quantum fluctuations.
If an engineer imbued with new ideas will able  to produce vacuum
energy density by means of quantum fluctuations, we imagine that
wormhole could be constructed physically.

 { \bf Acknowledgments }

          F.R is thankful to Jadavpur University and DST , Government of India for providing
          financial support. MK has been partially supported by
          UGC,
          Government of India under MRP scheme. \\

\pagebreak
\title{ \underline{\textbf{ Appendix }}: }

 \centering

 $ A =  \frac{1}{2}(f_1^{\prime\prime}+f_2^{\prime\prime})
     =  \frac{1}{2}[\frac{6Q^2}{a^4}-\frac{4M}{a^3}-Fu(u-1)a^{u-2}-Xw(w-1)a^{w-2}] $

$ B =  8\pi^2(\sigma^2-4a\sigma\sigma^\prime-2a^2\sigma^{\prime
        2})
    = \frac{1}{2a^2}(\sqrt{1-\frac{2M}{a}+\frac{Q^2}{a^2}}- \sqrt{1-F a^u - X a^w})^2
      +\frac{2}{a^2}(\sqrt{1-\frac{2M}{a}+\frac{Q^2}{a^2}}- \sqrt{1-F a^u -
      X
      a^w})(\frac{1-\frac{3M}{a}+\frac{2Q^2}{a^2}}{\sqrt{1-\frac{2M}{a}+\frac{Q^2}{a^2}}}
      -\frac{1-Fa^u-Xa^w+\frac{1}{2}Fua^u+\frac{1}{2}Xwa^w}{\sqrt{1-F a^u - X a^w}})
      - \frac{1}{a^2}(\frac{1-\frac{3M}{a}+\frac{2Q^2}{a^2}}{\sqrt{1-\frac{2M}{a}+\frac{Q^2}{a^2}}}
      -\frac{1-Fa^u-Xa^w+\frac{1}{2}Fua^u+\frac{1}{2}Xwa^w}{\sqrt{1-F a^u - X a^w}})^2     $

$  S  = \frac{(f_1^\prime-f_2^\prime)^2}{32\pi^2\sigma^2 a^2}
      =\frac{1}{2}\frac{(Fua^{u-1}+Xwa^{w-1}+\frac{2M}{a^2}-\frac{2Q^2}{a^3})^2}
     {(\sqrt{1-\frac{2M}{a}+\frac{Q^2}{a^2}}- \sqrt{1-F a^u - X a^w})^2}  $

$ T =
\frac{(f_1-f_2)(f_1^{\prime\prime}-f_2^{\prime\prime})}{32\pi^2\sigma^2
a^2} =
\frac{1}{2}\frac{(Fa^u+Xa^w-\frac{2M}{a}+\frac{Q^2}{a^2})[Fu(u-1)a^{u-2}+Xw(w-1)a^{w-2}-\frac{4M}{a^3}+\frac{6Q^2}{a^4}]}
     {[\sqrt{1-\frac{2M}{a}+\frac{Q^2}{a^2}}- \sqrt{1-F a^u - X a^w}]^2}        $

$ G =
 \frac{(f_1-f_2)(f_1^{\prime}-f_2^{\prime})}{16\pi^2\sigma^2
a^2} [ \frac{\sigma^{\prime}}{\sigma}+\frac{\sigma^{\prime}}{a}
+\frac{2}{a}] = \frac{(Fa^u+Xa^w-\frac{2M}{a}+\frac{Q^2}{a^2})
(Fua^{u-1}+Xwa^{w-1}+\frac{2M}{a^2}-\frac{2Q^2}{a^3})}
     {(\sqrt{1-\frac{2M}{a}+\frac{Q^2}{a^2}}- \sqrt{1-F a^u - X a^w})^2}
\linebreak   -\frac{1}{a}
\frac{1}{(\sqrt{1-\frac{2M}{a}+\frac{Q^2}{a^2}}- \sqrt{1-F a^u - X
a^w})}[(\frac{1-\frac{3M}{a}+\frac{2Q^2}{a^2}}{\sqrt{1-\frac{2M}{a}+\frac{Q^2}{a^2}}}
    -  \frac{1-Fa^u-Xa^w+\frac{1}{2}Fua^u+\frac{1}{2}Xwa^w}{\sqrt{1-F a^u - X a^w}})
    + \frac{1}{4\pi a^3}(\frac{1-\frac{3M}{a}+\frac{2Q^2}{a^2}}{\sqrt{1-\frac{2M}{a}+\frac{Q^2}{a^2}}}
    -  \frac{1-Fa^u-Xa^w+\frac{1}{2}Fua^u+\frac{1}{2}Xwa^w}{\sqrt{1-F a^u - X a^w}})+
    \frac{2}{a}]$
    $
C=\frac{2}{a^2}[(p+\sigma)+Y^\prime][\frac{(f_1-f_2)^2}{32\pi^2\sigma^3a^2}-8\pi^2\sigma
a^2] \linebreak = [ \frac{1}{8\pi a}(\sqrt{1-F a^u - X
     a^w}-\frac{1-\frac{3M}{a}+\frac{2Q^2}{a^2}}{\sqrt{1-\frac{2M}{a}+\frac{Q^2}{a^2}}})+\frac{1}{4\pi
     a^3}(\frac{Fua^u+Xwa^w}{\sqrt{1-F a^u - X a^w}})]-\frac{1}{8
     \pi a^2}[\frac{\sqrt{1-F a^u - X a^w}(Fu^2 a^{u-1}+Xw^2
     a^{w-1})+\frac{(Fu a^u+ Xw a^w)(Fu a^{u-1}+ Xw a^{w-1})}{2\sqrt{1-F a^u -
     X
     a^w}}}{1-F a^u - X a^w } ]\linebreak\times
     [\frac{4\pi}{a}[(\sqrt{1-\frac{2M}{a}+\frac{Q^2}{a^2}}- \sqrt{1-F a^u - X a^w})-\frac{(Fa^u+Xa^w-\frac{2M}{a}+\frac{Q^2}{a^2})^2}
     {(\sqrt{1-\frac{2M}{a}+\frac{Q^2}{a^2}}- \sqrt{1-F a^u - X a^w})^3}]]   $

$ H = \frac{(f_1-f_2)^2}{16\pi^2\sigma^2 a^2} [
\frac{2\sigma^{\prime}}{a \sigma}+\frac{{3\sigma^{\prime}}^2}{2
\sigma^2} +\frac{3}{2a^2}] =
\frac{(Fa^u+Xa^w-\frac{2M}{a}+\frac{Q^2}{a^2})^2}{(\sqrt{1-\frac{2M}{a}+\frac{Q^2}{a^2}}-
\sqrt{1-F a^u - X a^w})^2}[ \frac{3}{2a^2}- \frac{2}{a^2}
\frac{1}{\sqrt{1-\frac{2M}{a}+\frac{Q^2}{a^2}}- \sqrt{1-F a^u - X
a^w}}(\frac{1-\frac{3M}{a}+\frac{2Q^2}{a^2}}{\sqrt{1-\frac{2M}{a}+\frac{Q^2}{a^2}}}-
\frac{1-Fa^u-Xa^w+\frac{1}{2}Fua^u+\frac{1}{2}Xwa^w}{\sqrt{1-F a^u
- X a^w}})+
\frac{3}{2a^2}\frac{1}{\sqrt{1-\frac{2M}{a}+\frac{Q^2}{a^2}}-
\sqrt{1-F a^u - X
a^w}}(\frac{1-\frac{3M}{a}+\frac{2Q^2}{a^2}}{\sqrt{1-\frac{2M}{a}+\frac{Q^2}{a^2}}}-
\frac{1-Fa^u-Xa^w+\frac{1}{2}Fua^u+\frac{1}{2}Xwa^w}{\sqrt{1-F a^u
- X a^w}})^2]  $

$ N =    \frac{4\sigma^{\prime}}{a
^3}\frac{(f_1-f_2)^2}{32\pi^2\sigma^3 a^2}= \frac{1}{2\pi
a^5}[\frac{1-\frac{3M}{a}+\frac{2Q^2}{a^2}}{\sqrt{1-\frac{2M}{a}+\frac{Q^2}{a^2}}}-
\frac{1-Fa^u-Xa^w+\frac{1}{2}Fua^u+\frac{1}{2}Xwa^w}{\sqrt{1-F a^u
- X a^w}}][ \frac{(Fa^u+Xa^w-\frac{2M}{a}+\frac{Q^2}{a^2})^2 }
     {(\sqrt{1-\frac{2M}{a}+\frac{Q^2}{a^2}}- \sqrt{1-F a^u - X
     a^w})^3}]
     $
$ L =  \frac{32\pi^2\sigma \sigma^{\prime}}{a } =
-\frac{2}{a^4}(\sqrt{1-\frac{2M}{a}+\frac{Q^2}{a^2}}- \sqrt{1-F
a^u - X
a^w})[\frac{1-\frac{3M}{a}+\frac{2Q^2}{a^2}}{\sqrt{1-\frac{2M}{a}+\frac{Q^2}{a^2}}}-
\frac{1-Fa^u-Xa^w+\frac{1}{2}Fua^u+\frac{1}{2}Xwa^w}{\sqrt{1-F a^u
- X a^w}}] $


\end{document}